\documentclass[11pt]{article}

\columnwidth\textwidth

\begin{document}

\title{Comment on "$\!$A simple explanation of the non-appearance
of physical gluons and quarks$\,$"}

\author{Andreas Aste\\ 
Institut f\"ur Theoretische Physik der Universit\"at Basel,\\
Klingelbergstrasse 82, 4056 Basel\\
Switzerland}
\date{August 21, 2002}

\maketitle

\begin{abstract}
In a recent paper by Johan Hansson it is claimed that the
non-appearance of quarks and gluons as physical particles
is an automatic result of the nonabelian nature of the
color interaction in quantum chromodynamics.
It is shown that the arguments given by Hansson are insufficient to
support his claim by giving simple counter arguments.
\end{abstract}
\newpage

\section{Introduction}
The investigation of quark confinement has a long history in
theoretical physics. In \cite{hansson} it is claimed that
the non-appearance of gluons and quarks in the physical
spectrum is a trivial consequence of the nonabelian structure
of the corresponding Yang-Mills theory. It is the aim of this
comment to show that this assumption
is an oversimplification which can be rejected by very simple counter
arguments.

\section{Counter arguments}
It is a well known fact that confinement can also be observed
in the case of abelian gauge theories. The simplest example is
given by the Schwinger model \cite{schwinger,adam,aste}, i.e. quantum
electrodynamics in 1+1 dimensions with massless electrons and
photons. This model can be solved exactly and it turns out that
the fermions and the photon disappear from the physical spectrum.
What remains in the physical sector is a scalar particle which
has a mass that is proportional to the square of the coupling
constant.
The Schwinger model and other similar theories \cite{abdalla,schroer}
have served as toy models
in order to explain the confinement mechanism and
the infrared behaviour of quantum field theories.

In \cite{hansson} it is argued that confinement is due
to nonperturbative properties of QCD by considering the
equations of motion of the gluon field.
It is important to note that one should not mix up the two different
notions of {\em{classical}} and {\em{nonperturbative}}.
Classical solutions of the equations of motion of fields are
clearly nonperturbative, but this insight is only of restricted value
as far as the quantized theory is concerned.
Confinement is an intrinsically quantum field theoretical problem.

In \cite{hansson} it is argued that the representation of the
gluonic field with color index $b$
\begin{equation}
A_\mu^b(x)=\int \frac{d^3k}{2 \omega} \Bigl(a_\mu^b(\vec{k})e^{-ikx}
+a_\mu^b(\vec{k})e^{ikx} \Bigr), \quad b=1,..8
\end{equation}
does no longer hold in the nonabelian case, due to the nonlinear
structure of the equations of motion
\begin{equation}
(\delta_{ab}\partial^\mu+gf_{abc}A^\mu_c)(\partial_\mu A_\nu^b
-\partial_\nu A_\mu^b+gf_{bde}A_\mu^d A_\nu^e)=0. \label{nonlinear}
\end{equation}
But scattering theory constructs an $S$-matrix which relates
{\em{asymptotic}} initial and final states, where eq. (\ref{nonlinear})
does not cause any problems: The trivial plane wave solution
of a gluon field with 'fixed' color $a$
\begin{equation}
A_\mu^b(x)=\delta_{ab} \epsilon_\mu^be^{-ikx} \label{wave}
\end{equation}
solves the equation of motion, since $f_{abc}$ is totally
antisymmetric such that the gluon field with color index $b$
cannot couple to itself. Therefore we have no reason to assume
that plane wave solutions cannot be quantized in a canonical way
from a naive point of view.
Futhermore, additional "colored" but equivalent
solutions can be constructed from (\ref{wave}) by
gauge transformations.

I mention the fact that every solution of the Maxwell
equations for the photon field is also a solution of the corresponding
equations in the case of purely gluonic QCD, if the color of the
gluon field is held constant throughout space.
The confinement problem, which is still one of the most fundamental
problems in theoretical physics, must be tackled
by nonperturbative methods. One possibility is
lattice gauge theory, which allows the computation of correlation
functions in quantum field theory \cite{langfeld} by numerical
methods.

\section{Conclusions}

It is clear that confinement is most probably realized in
nature, and it is also well known that perturbative QCD is a
valuable tool only for the high energy regime of the theory.
Therefore, the simple Fock space structure which is used as
a basis of perturbation theory must lead to mathematical
inconsistencies.

Unfortunately, no proof of these statements is given in
\cite{hansson}. The classical equations of motion do not rule out
plane wave solutions in the case of purely gluonic QCD, and
arguments which go beyond the classical level are
missing in \cite{hansson}.

The assertion that the criterion for confinement is the
nonabelian structure of QCD (which leads to nonlinear equations
of motion) is already ruled out by the
observation that there {\em{are}} exactly solvable theories
which are abelian and which show the phenomenon of confinement.
Additionally, the equations of motion of coupled fields
(e.g. in quantum electrodynamics or the standard model) are
nonlinear, but we {\em{do}} observe asymptotic states in the experiment
which can be described by perturbation theory in a satisfactory
way.

\end{document}